\documentclass[oneside,a4paper,reqno]{amsart}
\usepackage{amssymb}

\theoremstyle{remark}

\begin{document}
\begin{titlepage}
\begin{center}
\bfseries  CONCERNING DICE AND DIVINITY
\end{center}
\vspace{1 cm}
\begin{center} D M APPLEBY
\end{center}
\begin{center} Department of Physics, Queen Mary, University of London, 
Mile End Rd, London E1 4NS, UK
 \end{center}
\vspace{0.5 cm}
\begin{center}
  (E-mail:  D.M.Appleby@qmul.ac.uk)
\end{center}
\vspace{0.75 cm}
\vspace{1.25 cm}
\begin{center}
\vspace{0.35 cm}
\parbox{12 cm }{Einstein initially objected to the probabilistic aspect of
quantum mechanics---the idea that God is playing at dice.  Later
he changed his ground, and focussed instead on the point that the
Copenhagen Interpretation leads to what Einstein saw as the abandonment of
physical realism.   We argue here that Einstein's initial
intuition was perfectly sound, and that it is precisely the fact that
quantum mechanics is a \emph{fundamentally} probabilistic theory which is
at the root of all the controversies regarding its interpretation. 
Probability is an intrinsically logical concept.  This means that the
quantum state has an essentially logical significance.  It  is extremely
difficult to reconcile that fact with Einstein's belief, that it is the
task of physics to give us a vision of the world apprehended
\emph{sub specie aeternitatis}.  
Quantum mechanics thus presents us
with a simple choice:  either to follow Einstein in looking for a theory
which is not probabilistic at the fundamental level, or else to
accept that physics does not in fact put us in the position of God
looking down on things from above.   There is a widespread fear that the
latter alternative must inevitably lead to a greatly impoverished,
positivistic view of physical theory.    It appears to us, however, that
the truth is just the opposite.    The Einsteinian vision  is much
less attractive than it seems at first sight.  In particular, it is
closely connected with philosophical reductionism. 
 }
\end{center}
\end{titlepage}
 For a long time Einstein strongly objected to the
indeterminism of quantum mechanics.  As he put it in a letter to
Born (written in 1926, in response to Born's proposal that the
wave-function has an essentially probabilistic significance): 

\begin{quotation}
Quantum mechanics is certainly imposing.  But an inner
voice tells me that it is not yet the real thing.  The theory says a lot,
but does not really bring us any closer to the secret of the `old one'. 
I, at any rate, am convinced that \emph{He} is not playing at dice.
[Born-Einstein letters~\cite{Born}, p.~91]
\end{quotation}
I think
people often find it difficult to understand why Einstein was so emphatic
in his rejection of a dice-playing God.  Quantum mechanics presents many
obstacles to the understanding.  But the concept of an objective chance
appears, on the face of it, intuitively very natural.  At least as judged
by the standards of commonsense, if anything is paradoxical, it is the
rigid determinism of classical physics, with its apparent denial of human
freedom.  It seems that Einstein came to feel this himself in the end. 
One finds him expressing strong objections  to the notion of a
dice-playing God as late as 1944 (ref.~\cite{Born}, p.149).  However in
1954 Pauli  reports him  as ``\emph{disput[ing]} that he
uses as criterion for the admissibility of a theory the question `Is it
rigorously deterministic?' ''(ref.~\cite{Born}, p.221).  

 It seems to me,
however, that Einstein gave in too easily.   It is precisely the fact
that quantum mechanics is a fundamentally probabilistic theory which is
at  the root of all the controversies regarding its interpretation. 
Specifically,  it is the fact that the wavefunction has a fundamentally
probabilistic significance which means that the wavefunction has to
collapse consequent on a measurement; and it is that collapse which makes
it hard to interpret the quantum state in the way that so many people
would like to interpret it, as a physically real entity.

 To illustrate
the point consider a case where someone---for the sake of definiteness let
us call her Alice---has bought a lottery ticket.  Suppose that the draw
has taken place, and that Alice's ticket won.  However, Alice does not
know this.  So, even though the reality is that Alice \emph{did} win the
lottery, she herself thinks that  the probability of her having won is
very small.  Now suppose that Alice opens a newspaper, and is surprised
to discover that hers is the winning ticket.  Then her state of mind will
suddenly change, from believing that the probability of her having won is
close to zero, to believing that it is close to one (not quite one
because the newspaper might have misprinted the winning number).  There
is, of course, nothing mysterious about this change in Alice's state of
mind.  Changing a probability assignment consequent on the acquisition of
new information is a very natural and reasonable thing to do.  

Now compare
this with a quantum mechanical measurement.  Before the measurement one
integrates the squared modulus of the electron's wave function and finds
that the probability of the electron having $x$ coordinate in the range
$1<x<2$ is $10^{-7}$.   But then one performs the measurement and finds
that the $x$ coordinate actually is in the range which one previously
considered to be highly improbable.  So, just as in the lottery example
considered in the last paragraph, one changes one's assessment of the
probability of the electron's $x$ coordinate being in the range $1<x<2$
from $10^{-7}$  to a value close to $1$.  Since the probability is
directly related to the wavefunction this means one must also make an
equally sudden and dramatic change to the electron's wavefunction.
That change is just the notorious collapse of the wavefunction, which I
think it is probably fair to say has been the cause of more philosophical
agonising than any other phenomenon in the history of physics. 

 The question is:  why should the
discontinuous change in the electron's wave function be considered any
more puzzling than the no less radical change in Alice's assessment of
the likelihood of her having won the lottery?  The short answer to this
question is that there is no temptation to regard the change in Alice's
beliefs  as anything more than a change in Alice's state of mind.   On
the other hand there is a very strong temptation to regard the electron's
wavefunction as a physically real entity.  Consequently, there is a
strong temptation to think that if the electron's wave function were to
change, merely as a consequence of the experimenter acquiring new
information, it would mean that reality itself had changed, merely as a
consequence of the experimenter acquiring new information.  However
physicists are, on the whole, reluctant to believe in spiritual
phenomena.  So the usual response is to try to twist the interpretation
of quantum mechanics in such a way as to make it seem, either that the
wavefunction does not \emph{really} collapse (as in, for example, the Bohm
or Everett interpretations), or else that it does collapse but not as a
consequence of the change in the experimenter's state of mind (as in
spontaneous collapse theories). 

At this point it will be convenient to introduce a piece of
terminology.  I will say that a probability is \emph{epistemic} if it is
conceived in the same way that one naturally conceives the probability of
Alice having won the lottery,  as representing, not a piece of
mind-independent physical reality, but only a cognitive agent's
expectations regarding that reality. Phrased in these terms the short
answer to the question posed at the beginning of the last paragraph
reads:  the discontinuous change in the probability of the electron being
located in a particular region tends to be seen as problematic because
there is a strong temptation to see the probability as, not merely
epistemic, but objectively real.   

However, the short answer is not completely
satisfactory because it fails to make clear why it is only with the
development of quantum mechanics that one finds this strong, almost
overwhelming temptation to regard probabilities as objectively real
entities.   After all, probability played a major role in the physics of
the nineteenth century.  Maxwell, in particular, was very clear that
probabilities are to be conceived as logical constructs, rather than
objective realities:
\begin{quotation}
They say that Understanding ought to work by the rules of right reason. 
These rules are, or ought to be, contained  in Logic; but the actual
science of Logic is conversant at present only with things either
certain, impossible, or entirely doubtful, none of which (fortunately) we
have to reason on.  Therefore the true logic for this world is the
calculus of probabilities, which takes account of the magnitude of the
probability which is, or ought to be, in a reasonable man's mind. 
[quoted Jeffreys~\cite{Jeffreys}, p.1]
\end{quotation}
This prompts the question:  if Maxwell, who was one of those chiefly
responsible for the classical theory of statistical mechanics, could
cheerfully accept that the probability of a molecule being in a
particular region is to be conceived in purely epistemic terms, why is it
that his $20^{\mathrm{th}}$ and $21^{\mathrm{st}}$ century  successors
are so wedded to the opposite, objectivist point of view?   Actually, it
is worth noting that Einstein himself was happy enough with probabilities
conceived in these Maxwellian, epistemic terms.  Indeed, some of
Einstein's most important contributions to theoretical physics relied on
a masterly deployment of
 probabilistic ideas.  It seems that Einstein was perfectly
willing to play dice \emph{himself}.  
His objection was only to the idea that
God might be playing them too.  The question is:  why?  Why should
behaviour which is acceptable in Einstein suddenly become unacceptable
when imputed to God?

I think the answer to that question must be that the theories of
classical physics were not \emph{fundamentally} probabilistic theories.
  To be
sure, the classical physicists found themselves compelled to use
probabilistic reasoning.  However, probabilistic concepts were not
embedded in their theories at the most fundamental level.  Maxwell could
easily accept that the probability distributions with which he worked
were not to be conceived as objective realities because there were
numerous other quantities in the theories of classical physics which
could be regarded as directly corresponding to physical realities.  The
problem his $20^{\mathrm{th}}$
and $21^{\mathrm{st}}$ successors face is that quantum
mechanics, by contrast, is probabilistic at the fundamental level.   In
quantum mechanics the only obvious candidate for a mind-independent
physical reality is the wave function.  So if one takes an epistemic view
of the wave function one seems to be driven into the position that
nothing at all in the theory corresponds to a mind-independent physical
reality.  It seems as though one is in danger of losing one's grip on
physical reality altogether.   And that, I think, is what leads to all
the attempts (the Bohm interpretation, the Everett interpretation,
spontaneous collapse theories, \emph{etc}.) to interpret the quantum state
as an objectively existent physical entity.

However, I think it is fair to say that these attempts are all highly
speculative.  Each such approach has its associated group of
enthusiasts.  However, the enthusiasts for one  approach
are unable to find  arguments  sufficient to persuade, either
the enthusiasts for any of the others, or the much larger
group of the conceptually uncommitted.   At least at present the decision
to opt for (say) the Bohm interpretation, rather than (say) the
Everettian, seems to rest on nothing more than personal taste. That
situation might conceivably change.  If, for example, Valentini's
hopes~\cite{ValentiniA,ValentiniB} were  fulfilled we would have solid
empirical reasons for preferring one objectivist interpretation over
another.  But in the present state of knowledge it is difficult to avoid
the suspicion that the question is empirically undecidable.  Of course,
``empirically undecidable'' does not mean ``necessarily false''.  Perhaps
the world is in reality Bohmian.  There is certainly nothing we know to
exclude that hypothesis.    The trouble is that there seems to be nothing
to support it either.  Not only are there no observations which would
clinch the question.  There does not even seem to be any moderately
persuasive reason for thinking the hypothesis likely.  

I intend no disrespect to the proponents of the various objectivist
interpretations by these remarks.  On the contrary it appears to me that
Bohm, Everett and others have made very important contributions to our
understanding of these questions.  Quantum mechanics is a deeply puzzling
subject, and I think that if one wants to understand it better  one needs
to look at it from every possible angle.  So I would certainly not dispute
the value of the work done by Bohm, Everett and others.   However, the
fact is that science is concerned specifically with those questions where
it is possible to find clearly stateable, cogent reasons for belief or
unbelief.  Until such reasons are forthcoming I do not see how any of the
objectivist interpreations can be considered a satisfactory solution to
the problem.  At the least I think it must be worth exploring alternative,
non-objectivist ways of thinking about the quantum state.

It is a striking fact that, although different people take very different 
views as to what the quantum state may be in ultimate reality, when  it
comes to the problem of making experimental predictions everyone
calculates in exactly the same way.  In particular, everyone collapses
the wave function (proponents of the Everett interpretation believe that
there is a state vector of the universe which does not collapse; however,
the wave function that an Everettian writes down on paper, for the
purposes of making an experimental prediction, collapses in just the same
way as the one that a Copenhagenist writes down).  At least so far as the
empirical predictions are concerned the significance of the quantum state
begins and ends with the fact that it specifies a set of probabilities. 
Of course, this does not logically exclude the hypothesis that the
quantum state has some other significance in ultimate reality. However,
in the absence of any compelling evidence as to what that significance
might be, it seems to me that the most natural and straightforward course
is to adopt the hypothesis that the quantum state simply is a compendium
of probablities, and to see what follows from that.  

If that is accepted the next question we have to address is, how to
interpret a probability statement.  
Here we run into the difficulty that the theory of probability is
troubled by a controversy which is even more long-standing than the
80-year old controversy about the interpretation of the wave function. 
At the beginning of the last century Poincar\'{e}~\cite{Poincare} (p.186)
described probability as an ``obscure instinct''.  In the 100 years that
have elapsed since then there has been much discussion.  However, the
effect has, if anything, only been to intensify the disagreements. 
Broadly speaking there are two schools of thought.  On the one hand there
is the objectivist school of thought (represented by, for example, von
Mises~\cite{MisesA,MisesB},  Fisher~\cite{Fisher} and
Popper~\cite{PopperA,PopperB}) which holds that a classical probability
distribution should be regarded as an objectively real physical entity,
which is what it is independently of anything that we might know or think
about it.  On the other hand there is the epistemic school of thought
(represented by, for example, Laplace~\cite{Laplace}, de
Finetti~\cite{Finetti}, Jeffreys~\cite{Jeffreys}, Savage~\cite{Savage} and
Jaynes~\cite{JaynesA,JaynesB}) which holds that a probability distribution
has an essentially logical significance. For a broad 
overview of the questions at issue see, for example,
Gillies~\cite{Gillies} and Howson and Urbach~\cite{Howson}.  For the
connection to quantum mechanics see Jaynes~\cite{JaynesA,JaynesB},
Fuchs~\cite{FuchsA,FuchsB}, Caves~\emph{et al}~\cite{CavesA,CavesB} and
Appleby~\cite{ApplebyC,ApplebyD}.

 Now one
might say that philosophy would no longer be philosophy if people ever
came to agree about something.  However, this particular dispute is
unlike many other conceptual disputes in that it has some immediate, and
very important practical consequences.  For, associated with the two
different schools of thought about the content of probability statements,
there are two very different statistical methodologies. The objectivist
view---the desire to interpret probabilities as mind-independent physical
entities which can be measured rather in the way that a mass can be
measured---motivated Fisher and others to develop what is now the
orthodox statistical methodology, described in every textbook.  By
contrast the epistemic point of view is associated with the statistical
methodology originally proposed by Bayes, and greatly extended by
Laplace.  These different statistical methodologies will, in general,
lead to different practical conclusions. 

The complaint of the Bayesians about the
orthodox statistical methodology has always been that it is (in the words
of de Finetti~\cite{Finetti}, p.245) ``\emph{ad hoc}'' and
``arbitrary''.   Jeffreys makes the point with characteristic irony when
he says of Fisher (one of the founding fathers of the orthodox
methodology)
\begin{quotation}
 I have in fact been struck repeatedly in my
own work, after being led on general principles to the solution of a
problem, to find that Fisher had already grasped the essentials by some
brilliant piece of common sense [Jeffreys~\cite{Jeffreys}, p.~393]
\end{quotation}
This is, in a way, a compliment. 
However, the compliment is distinctly back-handed:  for what Jeffreys is
really saying is that Fisher, notwithstanding his confusions and
inconsistencies, often contrives to get the right answer owing to the
power of his intuition.  It is rather as if a physicist were to
congratulate a snooker player on his ability to pot a ball
notwithstanding his  ignorance of Newtonian mechanics; or to
congratulate a fish on its ability to swim notwithstanding its 
ignorance of the principles of hydrodynamics.  I have argued
elsewhere~\cite{ApplebyC,ApplebyD} that that criticism is amply
justified.   Generally speaking what drives the Bayesian school of
thought is a desire for clarity and logical cogency. By contrast the
orthodox statistical methodology is driven by what Jaynes describes  as
an ideological conviction that, if statistics is to be scientific, then
probability distributions must be conceived as objectively real entities.
To attain that ideological end orthodox statisticians are willing to make
whatever sacrifice of logical coherence seems necessary.  

The problem orthodox
statisticians face is that, however sophisticated the technical
superstructure may  become, what is at the bottom of the pyramid is the
ordinary primitive intuition of one event being more or less probable
than another.  Furthermore every statistical argument has to rely on that
intuition if it is to make contact with reality.  One may cover hundreds
of pages with intricate calculations.   But the question one is
ultimately asking, and must answer if there is to be any point to the
calculations, is always very simple.  It is a question of the form:  how
probable is it that $X$?   Would it be wise to bet on $X$?  And
however words may be used by professional statisticians in their private
reasoning processes, the sense of the word ``probable'' as it is used in
the statement of the final conclusion is always the primitive sense,
which a child of $7$, who knows nothing of the formal apparatus of
probability theory, can comprehend.  In particular it is a sense of the
word ``probable''  which applies to single cases.  If I want to know
whether a drug is more likely to cure me than to kill me then, although I
might make use of data regarding the fate of other patients, the question
I want to answer is a single-case question:  what will probably happen to
\emph{me} if \emph{I} take the drug?  Shall I gamble with my life? 
(not the lives of a
statistical ensemble, but my singular,  personal life).

 Of
course the primitive notion of probability, that a child of $7$ can
understand, is non-quantitative.   It cannot be embedded in a  formal
mathematical theory without a very considerable degree of theoretical
elaboration.  The relation between the primitive notion of probability
and the formal mathematical one is, in some ways, similar to the relation
between the common sense notions of mass and force, and the concepts
going by the same names which feature in Newtonian mechanics.   So I am
certainly not meaning to \emph{identify} the primitive,
 common sense notion with
the formal mathematical concept of probability.  But what I think is
undeniably the case is that the formal concept is a development of the
primitive one (a very considerable development, no doubt, but a
development nonetheless).  Furthermore, the formal concept depends on the
primitive notion for its empirical applicability.  If the formal concept
is developed to the point where it loses all connection with the
primitive notion, then the theory will lose all its practical utility. 

   The problem this poses for orthodox statisticians (the
insuperable problem, as it seems to me) is that the primitive notion of
probability is, obviously and unavoidably, epistemic in its character. 
Consider the example I discussed earlier, where as it happens Alice's
lottery ticket won, but she does not know it.  In those circumstances
Alice believes that it is most unlikely that she won.  Most people would
consider she was \emph{right} to think it most unlikely.  
And yet the fact of
the matter is that she \emph{did} win.  It can be seen that we
 have here two
statements having radically different logical characters.  On the one
hand there is the epistemic statement: 
\begin{quotation}
Alice is most unlikely to have won
the lottery 
\end{quotation}
On the other hand there is the factual statement (which,
although Alice does not know it, describes the actual state of affairs)
\begin{quotation}
Alice did  win the lottery
\end{quotation}
 The second statement is a proposition about
the lottery draw, as it exists independently of what Alice knows or
does not know.  The first statement, by contrast, is as much about Alice,
and her limited information, as it is about the lottery.  Specifically, it
is a logically evaluative statement about what Alice, in her epistemic
situation, can reasonably expect.  This epistemic character becomes
apparent when one considers what happens when Alice opens the newspaper
and discovers that her ticket did win.  In that case her belief state
changes discontinuously.  Rather in the way that the wave function changes
discontinuously consequent on a measurement, Alice switches from thinking
that she almost certainly did not win the lottery to thinking that she
almost certainly did win (not quite certainly because, for example, the
newspaper might have misprinted the winning number or, for example,
because she might be hallucinating).  There is nothing mysterious, or
philosophically offensive about this discontinuous change.  It simply
reflects the fact that the statement was epistemic, and epistemic
statements naturally are subject to revision, consequent on the
acquisition of new information.  It is also worth noting that Alice's
discovery, that her ticket did win, will not lead her to think that she
was \emph{wrong} to believe that she probably had not won.   On the
contrary, she will continue to think that she was right to believe that
she probably had not won.   Naturally so:  for the belief that she
probably had not won did not represent a mistaken belief about the
physical world.  Rather it represented a logically evaluative belief
about what she, in her previous epistemic situation, could reasonably
expect.  

If this is accepted, and if the point I made earlier (that the
mathematical theory of probability relies on the primitive notion of
probability to make contact with reality) is also accepted, then it
follows that the project of the orthodox statisticians (to construct a
completely objective theory of statistical inference) is likely to run
into insuperable obstacles.  I believe that a more detailed
examination~\cite{Jeffreys,Finetti,JaynesA,JaynesB,ApplebyC,ApplebyD} of
the question confirms that proposition.   As a result the orthodox
statisticians are rather in the position that one would be in if one
insisted on trying to construct a theory of sound based on the assumption
that sound is a form of electromagnetic radiation; or a theory of number
based on the assumption that a number is, not an abstract logical entity,
but a concrete physical object.  It is only possible to produce a
simulacrum of success by bending the facts and twisting the logic.  

I believe that similar considerations apply to the quantum state. Quantum
mechanics is deeply and intriguingly different from
classical probability theory.  It is also much farther removed from
common sense ways of thinking.  Nevertheless it has certain basic
features in common.  In particular quantum probabilities, like classical
probabilities, are unavoidably epistemic in character:  as is 
clearly signalled by the discontinuous change in the quantum state which
occurs consequent on a measurement outcome.

This brings us back to the problem which troubled Einstein.  It is easy
to take an epistemic view of the probabilities in classical statistical
mechanics because classical statistical mechanics presents us with
various other mathematical constructs which can be thought of as
the depictions of actually existent physical entities.  But the same is
not true of quantum mechanics.  Consequently, it may seem that taking an
epistemic view of the quantum state amounts to giving up on the project
of understanding physical reality altogether.   It may look, on the face
of it, as though quantum mechanics thus interpreted leads,
 if  not to idealism, or to pure solipsism, then at any rate to a
depressingly positivistic  view in which physics begins and ends with the
task of predicting detector ``clicks''.   I believe it is that perception
which motivated Einstein's search for a more complete description, and
which continues to motivate the various  objectivist
interpretations of the quantum state.  

In response to that objection let me begin by saying that it is, to my
mind, a very reasonable objection.   Indeed, for most of my research
career my sympathies have been with the opponents of the Copenhagen
interpretation.  To a considerable extent they still are.  Bell complains
that the Copenhagen interpretation is ``unprofessionally vague and
ambiguous'' and that quantum mechanics, when interpreted in Copenhagen
terms, seems to be ``exclusively concerned with `results of measurement'
and [seems to have] nothing to say about anything else''
(Bell~\cite{Bell}, pp.~173 and 117).  I think he is right on both
counts.  I share Bell's conviction that the aim of physics is to
understand nature, and that counting detector ``clicks'' is not
intrinsically any more interesting than counting beans.  The day I become
convinced that physics does not in fact provide us with anything more
than procedures for  predicting detector ``clicks'' will be the day I
abandon physics in favour of some  more stimulating
activity.   Nevertheless I have gradually come to feel that the most
promising way forward lies, not in the Copenhagen Interpretation as such,
but in a greatly improved version of it to which Bell's criticisms would
not apply.  If I am asked to accept Bohr as the authoritative voice of
final truth then I cannot assent.  But if his writings are approached in
a more flexible spirit, as a source of insights which are but dimly
apprehended, then they suggest  a line of thought which I feel might, if
further developed, be very fruitful.

Taking this view does not amount to the abandonment of physical realism. 
If anything it is exactly the other way round.  The various
objectivist interpretations (Bohm, Everett, \dots) do indeed provide us
with numerous beguiling images of how things \emph{might},
\emph{conceivably} stand in ultimate reality.  The trouble is that there
does not seem to be any way to decide which, if any of these alternative
pictures corresponds to the truth.  As I stated earlier this situation
could change.  If, for example, Valentini's
hopes~\cite{ValentiniA,ValentiniB} were fulfilled we might have solid
empirical grounds for regarding the quantum state as a physically real
entity.  But in the present state of our knowledge objectivist
interpretations of the quantum state are purely speculative.  It is, of
course, true that every well-attested scientific theory started out as a
piece of  speculation.  However, although speculation undoubtedly has its
place in science, I think there is something very discouraging about a
theory which, one may reasonably fear, is never going to get beyond the
level of pure speculation.  The aim of science is, after all, not merely
to speculate, but to make empirically well-grounded statements about
physical reality.  It is not inconceivable that one or other of the
objectivist approaches will eventually be pushed to the point where it
meets that requirement.  However, my own feeling is that a more promising
approach is to try to make better sense of the wave function when it is
conceived epistemically.  Whether that judgment is correct only time will
tell.  But what is certain is that the motive is to better understand
what quantum mechanics is telling us about the world. 

It is, however, true that, although the epistemic approach does
not lead to the abandonment of physical realism, it forces us  to think
of physical reality in a very different way from the one to which we have
become classically accustomed.   For 200 years
after the time of Newton physics was inspired by the dream of
constructing a perfectly faithful depiction of the world. A map, each
point of which is in one-one correspondence with an element of physical
reality.  A map, furthermore, which leaves nothing out---a map in which
every feature of physical reality has its representative, and in which it
never happens that a simple feature of the map corresponds to a complex
feature of the actual world.  When Einstein talks of God I believe it is
this that he has in mind:  a vision of the world apprehended
\emph{sub specie aeternitatis}, as God might see it looking down on
things from above.     I think  it must be true that the epistemic
interpretation requires us to give up  on \emph{that}.   And I think it
must also be true that that to anyone of Einstein's philosophical
persuasion that is going to feel like a major sacrifice.  However, I would
suggest that if one takes the trouble to think through the implications
carefully one may start to feel that the sacrifice is not so great as it
initially looks.  In fact one may even begin to feel that it is not a
sacrifice at all, but a liberation.

The epistemic interpretation requires us to abandon the idea that a cat's
wave function is in one-one correspondence with the cat's ultimate
reality.  But that does not mean it requires us to give up the idea that
quantum mechanics tells us something important about the cat.  Quite the
contrary.  Quantum mechanics makes a large number of remarkably 
detailed statements about, for example, the cat's molecular biology.  The
epistemic interpretation leaves all of those statements completely
intact.  The statements are, to be sure, only probabilistic.  In other
words they are statements about what we, with our limited information,
can reasonably expect.  However, I would question whether that
acknowledgment represents the major intellectual defeat that Einstein
would take it to be.  It is not, after all, as though there has ever
\emph{really} been any question of apprehending the world \emph{sub specie
aeternitatis}.  God may know, with absolute certainty, the position of a
classical particle.  But we never have.  It is a truism that science does
not give us complete certainty.  The propositions of science are and
always have been, without exception, statements that something is more or
less probable.  Quantum mechanics, interpreted epistemically, changes
nothing in that respect.

If Einstein's hopes had been fulfilled we would be in the ostensibly
happy position of seeing right through to the cat's metaphysical bottom. 
At any rate, we would be able to imagine that we had achieved that
feat.  By contrast, the epistemic interpretation obliges us to concede to
the cat a degree of metaphysical privacy.  Is that such a bad thing? 
Einstein's vision, of the world apprehended \emph{sub specie
aeternitatis}, is closely connected
with philosophical reductionism.  And, however it may be with cats, the
idea that a human being simply reduces to a configuration of classical
fields (or whatever) is  not very plausible.  A large part of the
philosophy of the mind  consists of various rather unconvincing attempts
to understand how the brain, conceived in reductionist terms, can give
rise to consciousness.    One of the reasons I am interested in the 
epistemic point of view is that I feel that when properly developed it
may lead to a much more satisfactory, non-reductionist way of thinking
about the mind-brain relationship.

The ambition to ``know the mind of God'' is not realistic.  But I would go
further than that.  I would question whether the idea is even
attractive.  Suppose one really could comprehend the universe in its
entirety.  Might this not be found a little cramping?   If the universe
really could be comprehended in its entirety it would mean that the
universe was as limited as we are.  It seems to me that living in such a
universe would be rather like trying to swim in water that is only six
inches deep.  Groucho Marx once said that he would not want to belong
to a club that would have him as a member.  In a similar vein, my
personal feeling is that  I  would not wish to belong to a universe that I
was able to fully comprehend.

 Against this vision, of physics as knowing
the mind of God, I would like to set another:  physics as swimming in
water that is a great deal deeper than we are---perhaps even infinitely
deep.

\end{document}